\documentclass[aip,jcp,reprint,floatfix]{revtex4-1}

\usepackage[T1]{fontenc}
\usepackage[utf8x]{inputenc}

\pdfoutput=1

\usepackage[pdftex, bookmarks, pdffitwindow=false, pdfstartview=FitH, pdfdisplaydoctitle, colorlinks, plainpages=false,pdfauthor={Aleks Reinhardt}, pdftitle={Reentrant phase behaviour for systems with competition between phase separation and self-assembly}, pdfpagelabels, hypertexnames, pdflang={en}, hyperfootnotes=false, breaklinks]{hyperref}

\pdfpageattr {/Group << /S /Transparency /I true /CS /DeviceRGB>>}

\usepackage[dvipsnames]{xcolor}
\usepackage{tikz,graphicx,pgfplots,amsmath,amssymb}
\usetikzlibrary{calc}
\tikzstyle{every picture}+=[remember picture]
\pgfplotsset{
tick label style={
/pgf/number format/.cd,
fixed,
fixed zerofill,
precision=1,
}
}
\pgfplotsset{compat=newest}

\newcommand{\pdc}[3]{\ensuremath{\left( \frac{\partial #1}{\partial #2}\right)_{#3}}}

\tikzset{every plot/.style={prefix=plots/}}

\usepgfplotslibrary{external}
{paper-jcpstyle}
\tikzsetexternalprefix{figures/}

\usepackage{fixmath}

\usepackage{times}

\begin{document}

\title{Reentrant phase behaviour for systems with competition between phase separation and self-assembly}

\author{Aleks Reinhardt}
\altaffiliation{Contributed equally to this work.}
\author{Alexander J. Williamson}
\altaffiliation{Contributed equally to this work.}
\author{Jonathan P. K. Doye}
\email{jonathan.doye@chem.ox.ac.uk}

\affiliation{Physical and Theoretical Chemistry Laboratory, Department of Chemistry, University of Oxford, South Parks Road, Oxford OX1 3QZ, United Kingdom}

\author{Jes\'{u}s Carrete}
\author{Luis M. Varela}
\affiliation{Departamento de F\'{i}sica de la Materia Condensada, Facultad de F\'{i}sica, Universidad de Santiago de Compostela, E-15782 Santiago de Compostela, Spain}

\author{Ard A. Louis}
\affiliation{Rudolf Peierls Centre for Theoretical Physics, Department of Physics, University of Oxford,
1 Keble Road, Oxford OX1 3NP, United Kingdom}

\date{\today}

\begin{abstract}
In patchy particle systems where there is competition between the self-assembly of finite clusters and liquid-vapour phase separation, reentrant phase behaviour is observed, with the system passing from a monomeric vapour phase to a region of liquid-vapour phase coexistence and then to a vapour phase of clusters as the temperature is decreased at constant density.  Here, we present a classical statistical mechanical approach to the determination of the complete phase diagram of such a system.  We model the system as a van der Waals fluid, but one where the monomers can assemble into monodisperse clusters that have no attractive interactions with any of the other species. The resulting phase diagrams show a clear region of reentrance. However, for the most physically reasonable parameter values of the model, this behaviour is restricted to a certain range of density, with phase separation still persisting at high densities.
\end{abstract}


\maketitle 

\section{Introduction}
Self-assembly is a process by which a material, be it molecular, colloidal or macroscopic, spontaneously forms a well-defined structured aggregate.\cite{Whitesides2002}
There are many examples of such self-assembling systems, be they biological\cite{Goodsell2004} (\textit{e.g.}~virus capsids,\cite{Zlotnick2005} protein fibres\cite{Amos2004} and motors) or synthetic (\textit{e.g.}~surfactant micelles\cite{Antonietti2003} and supramolecular complexes\cite{Miras2010}), although the latter are not yet able to replicate the exquisite control of self-assembly exhibited by biological systems.
Indeed, some of the most remarkable synthetic self-assembling systems borrow from biology; for example, the use of DNA and RNA to form complex nanostructures\cite{Seeman2003} and devices.\cite{Bath2007}
From the materials perspective, the advantages of self-assembly are clear, potentially enabling new materials and devices to be fabricated, particularly those with features on the nanoscale.

For colloids and nanoparticles, there is a particular interest in developing particles that have `patchy' anisotropic interactions,\cite{Glotzer2007,Pawar2010} with a view to extending the range of structures that can be formed by self-assembly.
Experimental methods for synthesising such particles, although in their early stages, are advancing rapidly.\cite{Hong2006, Cho2007,Edwards2007,DeVries2007,Yang2008,Hong2008,Kraft2009,Kraft2009b,Wang2008,Mao2010,Sacanna2010}
Similarly, there is an increasing amount of computational work on patchy particles, both as simple models for understanding biological self-assembly,\cite{Hagan2006,Villar2009,Wilber2009b} and to explore what might be possible  with synthetic patchy particles and to learn the design principles for their successful self-assembly.\cite{Sear1999,Kern2003,Zhang2005,Bianchi2006,Wilber2007,Doye2007,Noya2007,Wilber2009,Sciortino2009,Romano2009,Romano2010,Noya2010,Sciortino2010}

One particular focus in the simulations of patchy particles has been the formation of finite aggregates, be they monodisperse\cite{Zhang2005,Wilber2007,Wilber2009} (similar to virus capsids) or polydisperse\cite{Sciortino2009,Miller2009,Sciortino2010,Whitelam2010} (micelle-like structures).
One intriguing result for these systems is that for certain parameter ranges, the competition between self-assembly and liquid-vapour phase separation can lead to `reentrant' phase behaviour.\cite{Wilber2007,Wilber2009,Sciortino2009,Sciortino2010}
A phase transition is said to be reentrant if it involves the transformation of a system from one state into a macroscopically similar (or identical) state via at least two phase transitions through the variation of a single thermodynamic parameter (such as the temperature).\cite{Narayanan1994}

\begin{figure}
\centering
\includegraphics{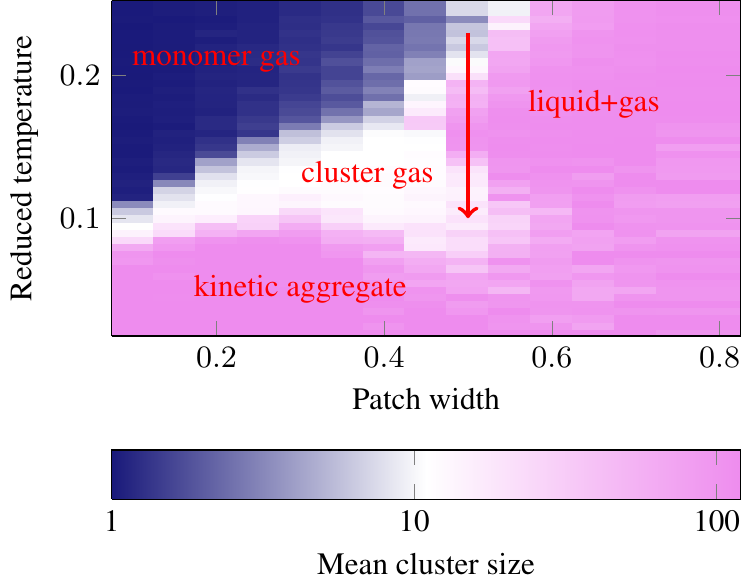}
\caption{The mean size of clusters formed as a function of the reduced temperature and the patch width for patchy particles designed to form 12-particle icosahedral clusters. These Monte Carlo simulations involved 120 particles simulated for $1.2\times10^8$ Monte Carlo steps  at a constant number density of $\rho\sigma^3=0.15$ and used the virtual move Monte Carlo algorithm.\cite{Whitelam2007} Results were averaged over 5 independent simulations. Decreasing the temperature at a patch width of approximately $0.5$ radians involves two phase transitions; from a monomer gas through a liquid-gas coexistence region to a cluster gas.}\label{fig-yieldplot}
\end{figure}

Such reentrant behaviour is illustrated in Fig.~\ref{fig-yieldplot} for a system that is designed to form monodisperse 12-particle icosahedra.\cite{Wilber2007} The figure shows the average size of clusters as a function of the temperature and the angular width of the patches at the end of dynamical simulations that initially started from a monomeric gas.
Values close to 12 indicate the formation of icosahedra, whereas very large cluster sizes indicate either a kinetic aggregate or the formation of a liquid droplet within a background vapour (and
thus correspond to a liquid-vapour coexistence region).
At very broad patch widths, the patches are relatively non-specific and the behaviour tends to that of a Lennard-Jones fluid with phase separation between a vapour of monomers and a liquid of monomers.
Conversely, at very narrow patch widths, the potential is so angularly dependent that the only thermodynamically stable phases are a vapour of monomers at higher temperatures and a vapour of clusters at lower temperatures (note that in these dynamical simulations, large aggregates form at low temperatures due to the slow kinetics, but this state is not a thermodynamically stable phase).
At intermediate patch widths, we expect competition between self-assembly and phase separation to occur.
For example, simulation results suggest that decreasing the temperature at a patch width of approximately 0.5 radians and at a constant density involves two phase transitions: first going from a \textit{gas} of monomers to a \textit{liquid-gas} mixture and then to a \textit{gas} of icosahedral clusters.
This reentrance is driven by the lower energy of the cluster gas phase (due to the strong internal bonding within the icosahedra) compared to the liquid phase, which has a greater entropy.

Clear evidence of reentrant phase behaviour has also been observed by Sciortino and co-workers in simulations of Janus particles with a single attractive patch. Certain parameter ranges lead to a competition between liquid-vapour phase separation and micelle formation.\cite{Sciortino2009,Sciortino2010}
In the phase diagrams that they computed, the standard behaviour of an ever-larger two-phase region in the $T$-$\rho$ phase diagram as the temperature is decreased below the critical point is perturbed by the formation of micelles, shrinking the coexistence region and shifting it to higher density.
However, what happens to this two-phase region at low temperature (does it end at a critical point, or persist to zero temperature?) is unclear, because the system instead formed a lamellar phase. In this paper, our aim is to address such questions by calculating complete phase diagrams for the fluid using a simple theoretical model of a system that can both self-assemble into clusters and exhibit liquid-vapour phase separation.

Reentrant phase behaviour is of course not limited to patchy particle systems. Indeed, the classic example of such behaviour is the nicotine-water mixture, which exhibits a closed loop in its temperature-composition phase diagram with both an upper and a lower critical solution temperature.\cite{Atkins}  Reducing the temperature whilst keeping the mole fraction of nicotine constant first leads to phase separation, and then to remixing driven by nicotine-water association. Liquid-liquid immiscibility leading to such phase diagrams is often attributed to hydrogen bonding.\cite{Davies1999,Jackson1991} Complete mixing will occur if the temperature is sufficiently high so as to overcome the enthalpic incompatibility of mixing. However, at sufficiently low temperatures, miscibility results from hydrogen bonding between different components in the system. Similar considerations have been used to explain reentrant phase behaviour seen in polymer aggregation in water,\cite{Nord1951} in non-ionic micellar solutions\cite{Corti1984} and in nematic reentrant phases in liquid crystals.\cite{Cladis1988}

Although the systems of interest in this paper are not two-component mixtures, unlike most systems where reentrant phase behaviour has previously been observed, the behaviour can nevertheless in some ways be analogous to such systems. The similarities arise due to the mixing of two states of the same component (monomers and associated clusters), rather than two components proper. An example of reentrant phase behaviour as a function of temperature in a system with a single effective density is ionic association.\cite{Levin1996,Fisher1993}
When dielectric constants are large, phase separation is driven by the usual enthalpic incompatibility. With low dielectric constants, coulombic-style phase separation between a low density (neutral cluster) structured gas phase and a high density (free ion) liquid phase occurs due to the long-range ionic interactions between the ionic components, rather than between the ions and the solvent. The formation of ionic clusters is somewhat similar to the cluster formation in the self-assembling patchy particle system that we focus upon, and so the reentrant behaviour (in simple theories of these ionic systems, `banana'-shaped coexistence curves are found\cite{Levin1996}) may have similar origins.

In this work, we investigate a theoretical model which can be used to
rationalise our earlier simulation results of self-assembling patchy colloidal
particles and one which yields a phase diagram reminiscent of the work of
Sciortino and co-workers.\cite{Sciortino2009,Sciortino2010} In particular, we are interested
in constructing a phase diagram for single-component systems that can
potentially form a vapour of monomers, a liquid of monomers and a gas of
monodisperse clusters. Here, these clusters are for the most part chosen to be
12-particle icosahedra,\cite{Wilber2007} but the approach can easily be
generalised to other clusters, as in subsection~\ref{subsect:results}. We have
chosen to model the chemical equilibrium between clusters and monomers with
explicit molecular partition functions, and to treat the vapour-liquid
equilibrium as a familiar van der Waals fluid; this is a more straightforward approach compared to using Wertheim theory,\cite{Wertheim1984,Wertheim1984b} which has already been used in the study of patchy particles.\cite{Bianchi2006}  We first look at just the monomer-cluster equilibrium in section~\ref{sec:mon-clus}, and then determine the necessary changes to the pure van der Waals fluid model in order to be able to form the three-state `self-assembling van der Waals fluid' and calculate appropriate phase diagrams for this system in section~\ref{sec:combined}.

\section{Monomer-cluster equilibrium}\label{sec:mon-clus}
\subsection{Partition functions}
We first construct the molecular partition function of an icosahedral cluster in a manner similar to that done previously for a simple model of tetrameric protein complexes.\cite{Villar-inpreparation,Villar2009}

The total molecular partition function of the icosahedral cluster is given by
\begin{equation}
\label{eq:qclus}
q_\text{clus} = q_\text{tr}q_\text{rot}q_\text{vib}\exp\left[ -E_\text{c}/kT  \right],
\end{equation}
where $E_\text{c}$ is the ground state energy of the cluster.
As we are comparing results to classical simulations, and since we are modelling colloidal particles, it is reasonable to use classical expressions for the
various partition functions.
The translational partition function is given by
\begin{equation}\begin{split} q_\text{tr} &=  \frac{V-BN_\text{clus}-bN_\text{mon}}{\Lambda^3} \\&= \frac{48 \sqrt{6}  ( \pi m k T)^{3/2}}{h^3}(V-BN_\text{clus}-bN_\text{mon}),\end{split}\end{equation}
where $\Lambda$ is the de Broglie thermal wavelength, the mass of the
icosahedron is $12 m$ (with $m$ being the mass of a monomer), and $N_\text{mon}$
and $N_\text{clus}$ refer to the number of monomers and the number of clusters
in the system, respectively. For the volume term, we choose to use the free volume approximation,\cite{Levin1996} which is one generalisation of the van der Waals expression to multiple components. The
volume available to clusters is thus $V-BN_\text{clus}-bN_\text{mon}$, taking into account the reduction in the total volume due to the presence of other clusters and monomers. Here, $B$ and $b$ are free volume coefficients (rather than hard-sphere second virial coefficients) for clusters and monomers, respectively.  The choice of values we assign to $B$ and $b$ is discussed in subsection~\ref{subsect-virial-coeffs}.

The moment of inertia about the principal axes of an icosahedron is given by $I_\text{c} = 80 m\sigma^2 / (5-\sqrt{5})$,
where $m$ is the monomer mass and $\sigma$ is the diameter of the monomers. The symmetry number of an icosahedral cluster is $60$. We can therefore calculate the rotational partition function as
\begin{equation}\begin{split} q_{\text{rot}} &= \frac{\sqrt{I_\text{c}^3 \pi}}{60} \times \left( \frac{8 \pi^2 k T}{h^2} \right)^{3/2} \\&= \frac{128\pi^{7/2}}{3}  \sqrt{1+\frac{2}{\sqrt{5}}}  \left(\frac{m k T \sigma^2}{h^2}\right)^{3/2}.\end{split}\end{equation}

There will be $(6\times 12-6)$ overall vibrational modes of the cluster. For simplicity, we assume that $(3\times 12 -6)$ probe the radial part of the potential, and $3\times 12$ probe the orientational degrees of freedom, and that all the modes of each type have the same spring constant.   Hence, we can write the vibrational partition function as
\begin{equation}
q_\text{vib} = \prod_i q_{\text{vib},\,i} \approx  \left(  \frac{kT}{\hbar \omega_\text{rad}} \right)^{30} \left(  \frac{kT}{\hbar \omega_\text{ang}} \right)^{36}  .
\end{equation}
To obtain expressions for $\omega_\text{rad}$ and $\omega_\text{ang}$, we need to assume some form for the interparticle potential.
We choose the potential that we used in our simulations of the self-assembly of icosahedra.\cite{Wilber2007,Wilber2009}
The potential is based on the Lennard-Jones form, but where the attractive region is modulated by an angular term that measures how well patches point at each other, and is given by
\begin{equation}
V_{ij}(\mathbf{r}_{ij},\,\mathbf{\Omega}_i,\,\mathbf{\Omega}_j)=\begin{cases}
V^{\text{LJ}}(r_{ij}) & r_{ij}<\sigma, \\
V^{\text{LJ}}(r_{ij})  V^{\text{ang}}(\hat{\mathbf{r}}_{ij},\,\mathbf{\Omega}_i,\,\mathbf{\Omega}_j) &  \sigma\le r_{ij} , \\
\end{cases}
\end{equation}
where $\mathbf{r}_{ij}$ is the inter-particle vector connecting the centres of the two particles $i$ and $j$, $r_{ij}$ is its magnitude,  $\mathbf{\Omega}_i$ and $\mathbf{\Omega}_j$ are the orientations of the particles $i$ and $j$, the Lennard-Jones potential is
\begin{equation}
V^{\text{LJ}}(r_{ij}) =
4 \varepsilon \left[\left(\frac{\sigma}{r_{ij}}\right)^{12}-\left(\frac{\sigma}{r_{ij}}\right)^6\right],
\end{equation}
and
\begin{equation}
V^{\text{ang}}(\hat{\mathbf{r}}_{ij},\,\mathbf{\Omega}_i,\,\mathbf{\Omega}_j) = \exp\left[\frac{-\theta_{k_\text{min}ij}^2}{2\,\sigma_\text{pw}^2}\right]\exp\left[\frac{-\theta_{l_\text{min}ji}^2}{2\,\sigma_\text{pw}^2}\right],
\end{equation}
where $\sigma_\text{pw}$ is the \textit{patch width} (we measure $\sigma_\text{pw}$ in radians), $\theta_{kij}$ is the angle between the patch vector of patch $k$ on particle $i$ and the interparticle vector $\mathbf{r}_{ij}$, and $k_\text{min}$ is the patch that minimises the magnitude of this angle. Hence, two particles interact only through a single pair of patches.

We assume that $\omega_\text{rad}$ and $\omega_\text{ang}$ are simply related to the second derivative of the radial and angular parts of the above anisotropic modified Lennard-Jones potential.  Namely, they are given by
\begin{equation}\omega_\text{ang}^2 =  - \frac{\varepsilon }{I_\text{mon}} \left( \frac{\partial^2 V^\text{ang}}{\partial \theta_{kij}^2}\right)_{ \left. \substack{ \theta_{kij} \\ \theta_{lji} }\right\} = 0}   =  \frac{\varepsilon }{I_\text{mon} \sigma_\text{pw}^2},\end{equation} where $I_\text{mon}$ is the moment of inertia of the monomer (given by $I_\text{mon} = (2/5) m \sigma^2$ for a spherical monomer), and
\begin{equation}\omega_\text{rad}^2  =  \frac{2}{m} \left( \frac{\partial^2 V^\text{LJ}}{\partial r^2}  \right)_{r=2^{1/6}\sigma}    = 72\times 2^{2/3} \times \frac{\varepsilon}{m \sigma^2}.\end{equation}
Although these approximations appear somewhat crude, this level of description is
reasonable given some of the other approximations used elsewhere. Furthermore,
using the exact form of the vibrational partition function would be unlikely to have a significant effect on the overall phase behaviour.

Finally, the Boltzmann factor in Eq.~\ref{eq:qclus} reflects the additional energy obtained as a result of clustering -- that is to say, the energy of bonding between the patches of the monomers when arranged in an icosahedron.  Ignoring next-neighbour interactions, the ground state energy of the icosahedron is $E_\text{c}=-30\varepsilon$, since each of the 12 monomers in an icosahedron is bonded to five other monomers with bond energy $\varepsilon$.

Overall, the icosahedral cluster molecular partition function is given by
\begin{equation}\begin{split}q_\text{clus} &= \left[ \frac{ k^{69} m^{36} T^{69} \sigma ^{69} \sigma _{\text{pw}}^{36} \exp\left[30 \varepsilon  / k T\right] }{h^{72} \varepsilon ^{33}}\right] \times \\  & \qquad \times \left[V-B N_\text{clus} - b N_\text{mon}\right] \times 9.4\times 10^{20}. \end{split}\label{qclus}\end{equation}

The monomers have both rotational and translational degrees of freedom, but not generic attractions at this stage. Therefore, the molecular partition function of the monomer is given by  $q_\text{mon} = q_\text{tr, mon}\times q_\text{rot, mon}$, where
\begin{equation}
q_\text{tr, mon} = \frac{V-B N_\text{clus}-bN_\text{mon}}{\Lambda^3}
\end{equation}
and
\begin{equation} q_\text{rot, mon} = \frac{16 \sqrt{2} \pi^{7/2} \left(I_\text{mon} k T\right)^{3/2}}{ sh^3 }  = \frac{64 \pi ^{7/2} \sigma ^3 (m k T)^{3/2}}{25 \sqrt{5} h^3}.\label{eqn-qrotmon}\end{equation}
Note that the symmetry number $s$ of a monomer is 5, since there are five identical patches on each monomer, and the same volume term occurs as in the cluster translational partition function because we are using the free volume approximation.\cite{Levin1996} The overall monomer molecular partition function is therefore given by
\begin{equation} q_\text{mon} = \frac{128 \sqrt{\frac{2}{5}} \pi^5 k^3 m^3 \sigma^3 T^3}{25 h^6}\times \left[V-B N_\text{clus} - b N_\text{mon}\right].\end{equation}

\subsection{Estimation of free volume coefficients}\label{subsect-virial-coeffs}
We choose to use the free volume approximation\cite{Levin1996} summation to take into account the effects of excluded volume. Although there are more accurate functions which describe such effects better,\cite{Laghaei2006} we deliberately wish to keep the partition functions as simple as possible, as we are interested in the \emph{generic} properties and phase behaviour resulting from the competition between self-assembly and phase separation, rather than the detailed behaviour of a specific system. Moreover, we want a form consistent with the van der Waals description of the fluid, and one that works reasonably well both at high and, to a degree, at low system densities.

Within the free volume approximation,\cite{Levin1996} the maximum density in the system will equal the reciprocal of the free volume coefficient $b$, $\rho_\text{max}=1/b$. Throughout this paper, all densities $\rho$ refer to number densities, $\rho=N/V$, rather than mass densities. We choose to set the maximum density achievable in the system to $1\,\sigma^{-3}$, and hence $b=\sigma^3$.
This choice is reasonable both in terms of the comparison of the maximum
packing fraction with that for random close packing, and of the position of
the van der Waals critical point with that of the Lennard-Jones fluid.

The relative values of $B$ and $b$ should be proportional to the cluster and monomer sizes. To determine the size of the cluster relative to that of the monomer, we assume that clusters are spherical objects, but such that we take their icosahedral character into account. The circumradius of an icosahedron\cite{CRC-tables} with edge length $a$ is $r = (a/4) \sqrt{10 +2\sqrt{5}} \approx 0.95\,a$; if the icosahedron is built of monomeric spheres of diameter $\sigma$, then $a=\sigma$. However, we also need to take into account the fact that these monomers protrude out of the icosahedron itself, and so another hard-core radius of these constituent spheres, $\sigma/2$, must be added to the circumradius of the icosahedron to estimate the overall cluster radius $R$, giving $R=1.45\,\sigma$.

The volume occupied by this sphere relative to the volume occupied by a single
particle is $R^3 / (\sigma/2)^3 = 24.4$. This is therefore a suitable measure
of how much bigger the icosahedral cluster is compared to a single patchy
particle, and suggests that an appropriate ratio of cluster and monomer
coefficients is $B/b\approx 24$. We consider this case most comprehensively, and also a few other values in order to characterise more fully the types of phase diagram that may arise. This ratio of $B/b$ is specific to the icosahedral system described, and would take values different from the ones estimated here if the system were allowed, for example, to adopt a different cluster shape (for the same number of monomers) or to assemble into a different oligomeric state.

\subsection{Clustering transition}\label{subsect:equiline}
We assume that the only relevant reaction in the system is $12\mathrm{A} \rightleftharpoons \mathrm{A}_{12}$, which is to say that there are no partially constructed clusters whatsoever. We justify this assumption from simulations, which show that the free energy of a 12-particle cluster is considerably lower than any intermediate size bigger than a monomer.\cite{Wilber2007}  At equilibrium, the chemical potentials of the individual species, appropriately weighted by their stoichiometric coefficients, will be equal; in this case, $12\mu_\text{mon}=\mu_\text{clus}$, where $\mu_i = -kT\ln (q_i/N_i)$. We choose to define the centre of this equilibrium as the point at which the probability of a particle being in a cluster or a monomer is equal, which imposes the condition $\rho_\text{mon}=12\rho_\text{clus}$, where $\rho_\text{mon}=N_\text{mon}/V$ and $\rho_\text{clus}=N_\text{clus}/V$. We denote the temperature at which both conditions are fulfilled as the clustering temperature, $T_\text{clus}$.

\begin{figure}[tbp]
\centering
\includegraphics{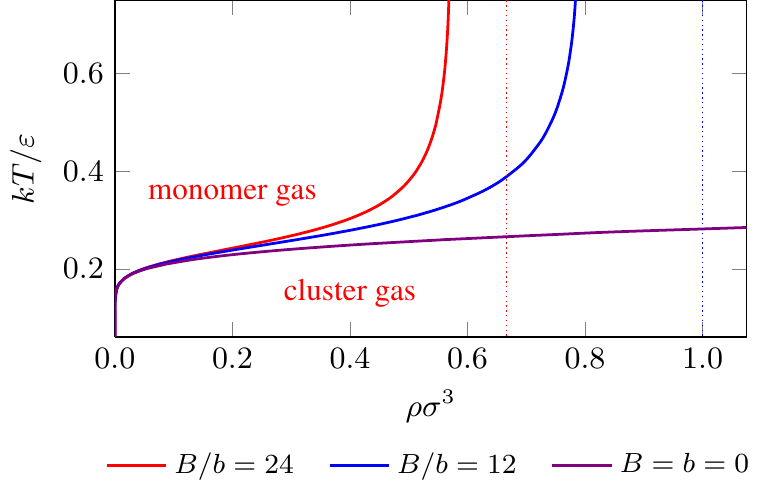}
\caption{$T_\text{clus}(\rho)$ for the transition from monomers to icosahedral clusters for different ratios of the free volume coefficients, $B/b$. For the non-ideal systems, $b/\sigma^3=1$, and dotted lines show $\rho_\text{max}^\text{eql}\sigma^3$, the maximum density at which particles can have an equal probability of being monomeric or in a cluster.
$\sigma_\text{pw}=0.5$.\label{fig-cluster-gas} }
\end{figure}

\begin{figure}[tbp]
\centering
\includegraphics{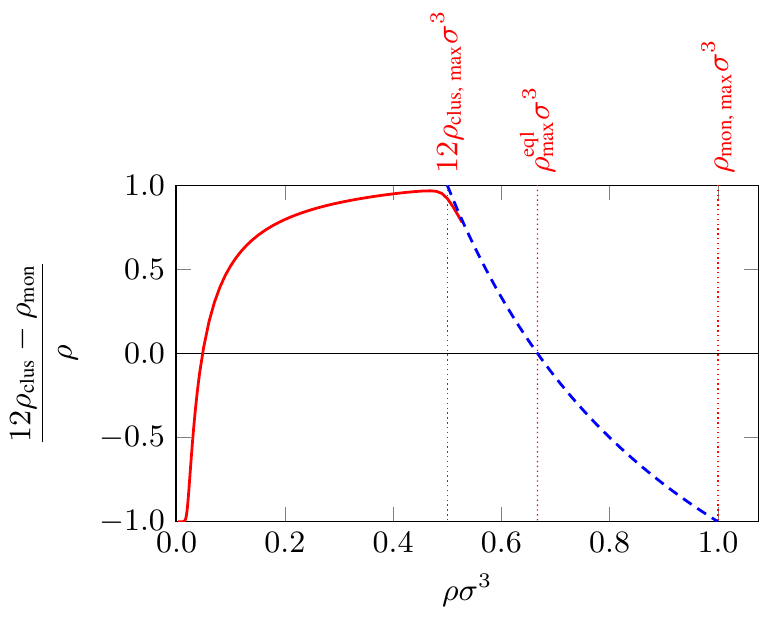}
\caption{The relative proportion of monomers and clusters in the equilibrium
  mixture is shown as a function of the density at $kT/\varepsilon=0.2$ for $B/b=24$, $\sigma_\text{pw}=0.5$. The function $(12\rho_\text{clus}-\rho_\text{mon})/\rho$ takes the value $-1$ when only monomers are present, $+1$ when only clusters are in the system, and 0 at the centre of the clustering transition (\textit{i.e.}~on the $T_\text{clus}(\rho)$ line). Vertical dotted lines indicate the maximum particle density associated with clusters $12\rho_\text{clus,\ max}$, the maximum density at which the monomers and clusters can have an equal probability $\rho_\text{max}^\text{eql}$ and the maximum monomer density $\rho_\text{mon,\ max}$, as labelled. The solid line is calculated from the partition functions, whilst the dashed line is the extrapolation in the regime where the available volume is held at zero.\label{fig-cluster-gas-cutThrough} }
\end{figure}

We are not able to solve this system of equations analytically, and so we use numerical minimisation of the difference in chemical potential using the density constraint to obtain $T_\text{clus}$ as a function of density, as plotted in Fig.~\ref{fig-cluster-gas}. The transformation between the two states is continuous (see Fig.~\ref{fig-cluster-gas-cutThrough}) and the $T_\text{clus}(\rho)$ line is not a phase coexistence curve, but simply the centre of a chemical equilibrium that indicates where clusters and monomers are equally probable states.

We first consider the ideal limit ($b=0$ and $B=0$) and see from Fig.~\ref{fig-cluster-gas} that at low temperatures, the cluster gas is more stable than the monomer gas.
Although the clusters have a lower entropy as a consequence of the lower number of translational degrees of freedom compared to the monomer gas, at sufficiently low temperatures, the clusters will have a lower free energy, because their lower entropy is more than overcome by their lower energy due to bonding between the patches.
Also, the equilibrium favours the cluster state as the density increases (\textit{i.e.}~$T_\text{clus}$ increases with $\rho$), since the effective number of particles in the gas phase is lower.

Although the above are the most important basic features of the monomer-cluster equilibrium, there are a few more features that we should consider. For example, the model predicts that at high temperatures, the cluster gas becomes favoured. This behaviour is an unphysical consequence of applying the harmonic approximation in a temperature range where one expects all bonds to be dissociated. However, this deficiency in the model is not a problem for the current study, as we are interested in temperatures that are well below where this phenomenon occurs.

Figure~\ref{fig-cluster-gas} also shows the effect of introducing excluded volume interactions on the monomer-cluster equilibrium. There is no maximum density in principle for the ideal system ($b=0$, $B=0$), and clusters are always stable to the right of the clustering transition line. The first effect of the excluded volume is to limit the maximum density. For the $B/b=24$ case, the maximum density of monomers is $\rho_\text{mon,\ max}\sigma^3=1$, and the maximum particle density associated with the clusters is $12\rho_\text{clus,\ max}\sigma^3=1/2$. For the $B/b=12$ case, both the monomers and the clusters have a maximum particle density of $1$.

At low densities, the excluded volume interactions have little effect and the $T_\text{clus}(\rho)$ lines initially follow that for the ideal case. However, as the density increases, the lines are displaced upwards with respect to the ideal case because the volume term $V-BN_\text{clus}-bN_\text{mon}$ in the translational partition functions of both monomers and clusters decreases, and this destabilises the monomeric state more because it has a greater number of translational degrees of freedom.

A typical dependence of the fluid composition on the density is shown in Fig.~\ref{fig-cluster-gas-cutThrough} for a system in which $B/b>12$. The fraction of particles in clusters increases rapidly as the system passes through the clustering transition on increasing the density, and as $\rho=12\rho_\text{clus,\ max}$ is approached, this fraction is usually close to unity. However, when clusters exclude more volume than 12 monomers (which is the usual physical situation), \textit{i.e.}~when $B>12b$, achieving densities beyond $12\rho_\text{clus,\ max}$ is only possible if the monomeric state again becomes populated, because monomers pack more efficiently (for example, when $B/b=24$, 12 monomers take up only half the volume of a single icosahedral cluster). Hence we see in Fig.~\ref{fig-cluster-gas-cutThrough} that as the density increases further, the fraction of monomers increases, becoming the dominant species beyond $\rho_\text{max}^\text{eql}$, the maximum density for which particles can have an equal probability of being monomeric or in a cluster.\footnote{$\rho_\text{max}^\text{eql}=24/(B+12b)$, where this expression is the simultaneous solution of both $V-BN_\text{clus}-bN_\text{mon}=0$ and $12\rho_\text{clus}=\rho_\text{mon}$. For $B/b=24$, $\rho_\text{max}^\text{eql}\sigma^3 = 2/3$; for $B/b=12$, $\rho_\text{max}^\text{eql}\sigma^3 = 1$.} In Fig.~\ref{fig-cluster-gas}, the dotted vertical lines correspond to $\rho_\text{max}^\text{eql}$, and so for $B/b=24$, monomers dominate to the right of the line. We also note that in practice, it becomes numerically extremely difficult to calculate the equilibrium densities of monomers and clusters beyond $\rho=12 \rho_\text{clus,\ max}$, because the $V-BN_\text{clus}-bN_\text{mon}$ term in the translational partition function becomes very close to zero and the chemical potentials of the clusters and the monomers both become extremely large. However, it is relatively easy to extrapolate beyond $12 \rho_\text{clus,\ max}$ using the approximation that the available volume $V-BN_\text{clus}-bN_\text{mon}$ is zero, as illustrated in Fig.~\ref{fig-cluster-gas-cutThrough}.

We note here that when considering the combined self-assembling van der Waals fluid in section~\ref{sec:combined}, for the most part we need not worry about some of these phenomena that occur when $\rho>12\rho_\text{clus,\ max}$, because phase separation intervenes before this density is reached.

\section{Self-assembling van der Waals fluid}\label{sec:combined}
Having now developed an approach to calculate the equilibrium between a vapour
of monomers and clusters, we now wish to calculate the phase diagram of a `self-assembling' van der Waals fluid, where monomers can either assemble into inert clusters (with no attractive interactions) or condense to form a liquid. We have chosen to use a van der Waals fluid as a starting point because, although we could use a more sophisticated description of the monomer and liquid states, we are mainly interested in the underlying fundamental physical behaviour of the system rather than in trying to describe the behaviour of any one particular experimental system.

\subsection{Partition functions}
To model clusters, we use the cluster molecular partition function from Eq.~(\ref{qclus}) above. The canonical partition function for a van der Waals (vdW) monatomic fluid is given by
\begin{equation} Q=\frac{1}{N!}\left( \frac{2\pi m k T}{h^2}  \right)^{(3N/2)} (V-Nb)^N \exp\left[ \frac{aN^2}{VkT} \right], \label{eqn-Qvdw} \end{equation}
where $a$ and $b$ are positive constants.\cite{McQuarrie2000} This expression is analogous to that for an ideal gas, but with an excluded volume term ($Nb$) and an attractive exponential term.
To model the vapour and the liquid, we use the molecular partition function for a van der Waals fluid derived from Eq.~(\ref{eqn-Qvdw}), except for two modifications. First, we include an additional term in the volume expression in order to take into account the exclusion of monomers by clusters as well as by monomers themselves. Using the free volume approximation\cite{Levin1996} leads to the same available volume term as in Eq.~(\ref{qclus}), namely $V-BN_\text{clus}-bN_\text{mon}$. Furthermore, we multiply this molecular partition function by $q_\text{rot}$ of the same form as in Eq.~(\ref{eqn-qrotmon}) to take into account the fact that the monomers are now particles with orientational degrees of freedom. There are no attractive interactions between clusters, or between clusters and monomers, to reflect the fact that in our patchy particle model, no patches are exposed on the surface of the clusters (we ignore partially formed clusters); all patches are instead used in the internal bonding.

\begin{figure*}[tbp]
\centering
\includegraphics{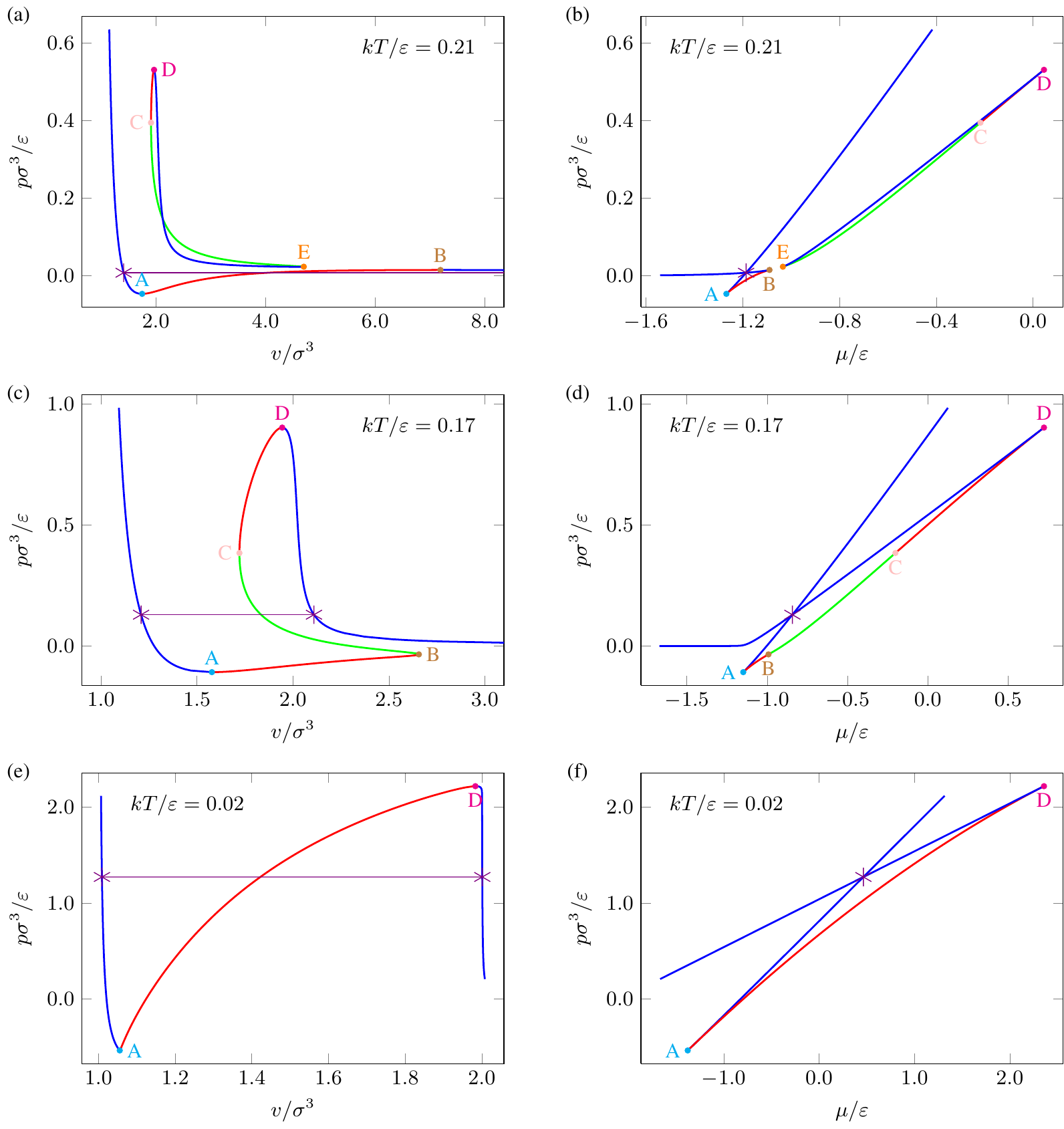}
\caption{The $p$-$v$ ((a), (c) and (e)) and $p$-$\mu$ ((b), (d) and (f)) curves for the self-assembling van der Waals fluid at reduced temperatures of (a) and (b) $kT/\varepsilon=0.21$, (c) and (d) $kT/\varepsilon=0.17$  and (e) and (f) $kT/\varepsilon=0.02$.  Regions of mechanical stability are coloured blue, those of mechanical instability red, and the mechanically stable, but compositionally unstable, region of back-bending is coloured green. Binodal points are marked by violet asterisks connected by a tie line. Spinodal points, where the derivative of the pressure with respect to the volume is either zero or infinity, are also shown with the labelling and colour-coding matching that used in Fig.~\ref{fig-overall-phase-diag}. $v=V/N$, $a/\varepsilon\sigma^3=1$, $b/\sigma^3=1$, $\sigma_\text{pw}=0.5$, $B/b=24$.}\label{fig-PVdiags}
\end{figure*}

\subsection{Coexistence curve}
The coexistence curve, \textit{i.e.}~the curve where the difference in the Helmholtz energy between two phases is zero, is often calculated using a Maxwell construction, whereby a straight line is constructed in the $p$-$V$ phase diagram over the mechanically unstable region such that the areas of the regions defined by the van der Waals loop and this line are equal.  However, in many interesting cases, van der Waals loops do not exist,\cite{Levin1996,Fisher1993} and it would be useful to have a more generic (and, indeed, easier) approach. We can calculate the chemical potential of species $j$ using $\mu_j = (\partial A/\partial N_j)_{V,\,T,\,N^\prime}$,
where $N^\prime$ refers to all components other than $j$, and the pressure as $p=-(\partial A/\partial V)_{T,\,N}$. Plotting the chemical potential against the pressure parametrically allows us to determine binodal points where the curves cross. Whenever the chemical potential defines two (or more) different pressures, the stable phase invariably has the \emph{largest pressure}, which maximises the entropy and is consequently favoured by the second law of thermodynamics.\cite{Levin1996,Hillert2008}

As in section~\ref{sec:mon-clus}, we assume that the only relevant reaction in the system is $12\mathrm{A} \rightleftharpoons \mathrm{A}_{12}$. We then seek to find the coexistence curves for the system where van der Waals monomers and clusters are in equilibrium. This is achieved by setting $\mu_\text{clus}=12\mu_\text{mon}$ and constructing $p$-$\mu$ curves to calculate binodals. Since the density of clusters is a function of the density of van der Waals monomers, and vice versa, we use numerical techniques for solving the system of equations. To find $p$ as a function of $\mu$, we impose a chemical potential $\mu$ and numerically minimise both $\left|\mu_\text{clus}-12\mu_\text{mon}\right|$ and $\left|\mu_\text{clus}/12+\mu_\text{mon}-2\mu\right|$ as functions of $\rho_\text{mon}$ and $\rho_\text{clus}$, all under the constraints that densities are positive and that the available volume ($V-BN_\text{clus}-bN_\text{clus}$) is non-negative. The first minimisation ensures that the system is at equilibrium ($\mu_\text{clus}\approx 12\mu_\text{mon}$), whereas the second fixes the individual chemical potentials to the imposed value ($\mu_\text{clus}\approx 12\mu$ and $\mu_\text{mon}\approx \mu$). We only admit answers which have a combined difference equalling zero to within a small error term of $10^{-6}$, although in practice the error term is generally several orders of magnitude smaller.  Having now calculated $\rho_\text{clus}$ and $\rho_\text{mon}$, we can calculate the pressure. Example $p$-$v$ and $p$-$\mu$ curves are given in Fig.~\ref{fig-PVdiags}. By performing this analysis at a large number of relevant temperatures, the full phase diagram can be calculated.

\subsection{Stability criteria}\label{subsect:stabcrit}
To include spinodals on the phase diagram, we can bracket regions of
mechanical stability by finding where the gradients of the $p$-$v$ curves change sign. Notice from Fig.~\ref{fig-PVdiags}(c) that there are parameter ranges where there are two regions of mechanical instability (AB and CD), separated by a region of mechanical stability (BC).
In this system, however, we also have to check for compositional stability, \textit{i.e.}~whether fluctuations away from the equilibrium composition of monomers and clusters are stable.
To derive a condition for compositional stability, we start from the fundamental equation for the Helmholtz energy, which may be written as
\begin{equation}
\mathrm{d} A = {}-{}p\,\mathrm{d} V - S\,\mathrm{d} T + \mu_\text{mon}\,\mathrm{d} N_\text{mon} + \mu_\text{clus}\,\mathrm{d} N_\text{clus}.
\end{equation}
However, the number of monomers and clusters is not independent, but
$N=12 N_\text{clus} + N_\text{mon}$ and $\mathrm{d} N = 12\,\mathrm{d} N_\text{clus} + \mathrm{d} N_\text{mon}$. Hence, we may rewrite the above equation as
\begin{equation}\begin{split}
\mathrm{d} A & = {}-{}p\,\mathrm{d} V - S\,\mathrm{d} T + \mu_\text{mon}\,\mathrm{d} N  {}+{} \\
 & \qquad\qquad {}+{} \left(  \mu_\text{clus}-12\mu_\text{mon} \right)\,\mathrm{d} N_\text{clus}.\end{split}
\end{equation}
It follows that
\begin{equation}
\pdc{A}{N_\text{clus}}{N,\,V,\,T} = \mu_\text{clus}-12\mu_\text{mon},
\end{equation}
which is of course zero at equilibrium. The condition for compositional stability is hence that
\begin{equation}
 \left(\frac{\partial^2 A}{\partial N_\text{clus}^2}\right) = \pdc{(\mu_\text{clus}-12\mu_\text{mon})}{N_\text{clus}}{N,\,V,\,T} > 0.\label{eqn-comp-stab}
\end{equation}
This stability criterion can be evaluated at every equilibrium
point of the combined system. The system is compositionally unstable exactly along the line BC.

\begin{figure*}[tbp]
\centering
\includegraphics{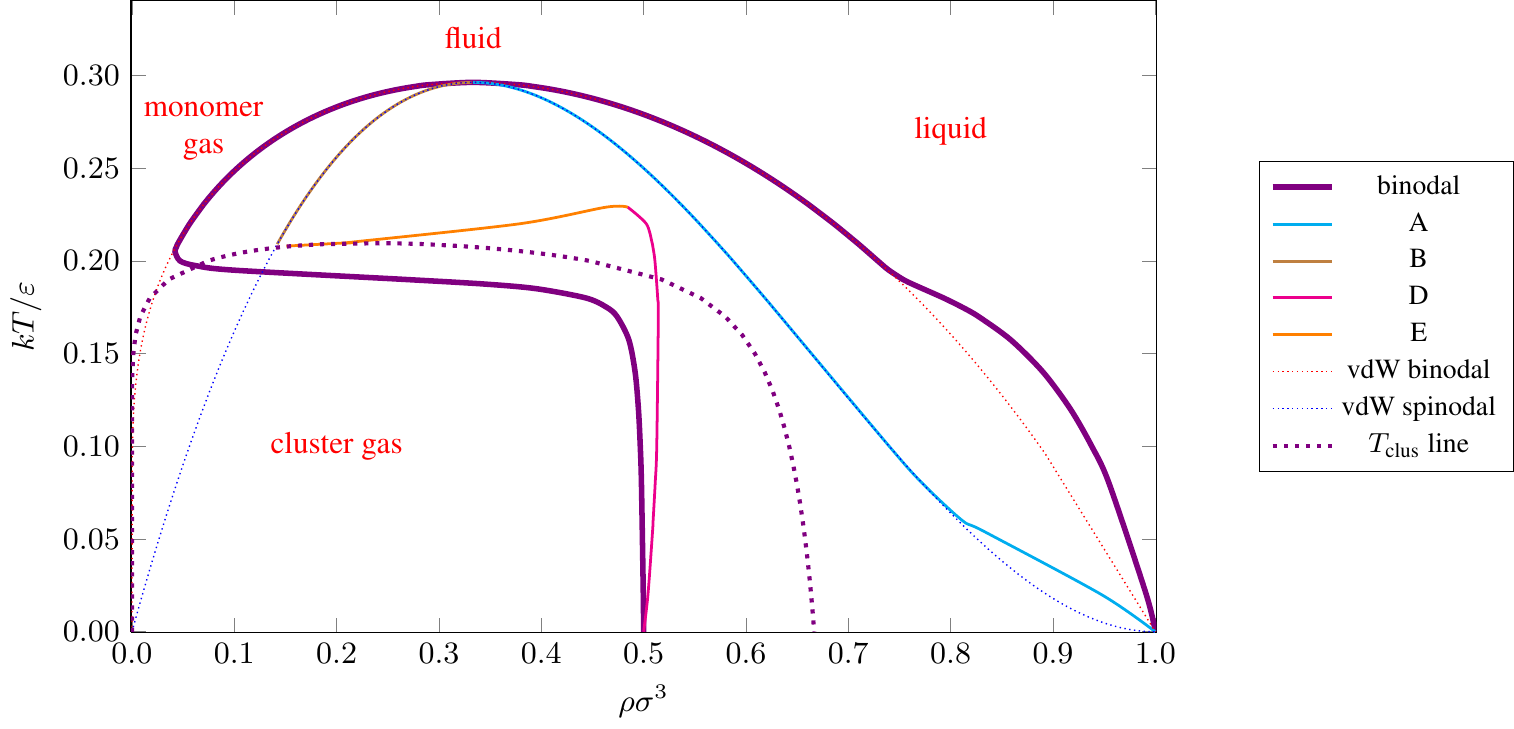}
\caption{The $T$-$\rho$ phase diagram for the self-assembling van der Waals fluid for $B/b=24$. The thick lines give the binodals for the system. Also represented are lines which correspond to points A, B, D and E in the $p$-$v$ plots in Fig.~\ref{fig-PVdiags}, where the derivative of the pressure with respect to the volume is zero or $\pm\infty$, where they are true limits of stability. The binodals and spinodals for the standard, pure van der Waals fluid are plotted for comparison (dotted lines). Finally, the clustering temperature ($T_\text{clus}$) is also plotted (thick dotted line).  $a/\varepsilon\sigma^3=1$, $b/\sigma^3=1$, $\sigma_\text{pw}=0.5$.\label{fig-overall-phase-diag}}
\end{figure*}

\subsection{Phase diagrams}\label{subsect:results}
We are most interested in computing phase diagrams for our system in the region of parameter space where reentrant behaviour might be expected to occur. We choose a value for the patch width from the region where simulations suggest reentrant phase behaviour occurs (Fig.~\ref{fig-yieldplot}), namely $\sigma_\text{pw}=0.5$. We choose the free volume coefficients $B$ and $b$ in line with subsection~\ref{subsect-virial-coeffs}, such that $b/\sigma^3=1$, and use the case $B/b=24$ as our main example. We then choose the van der Waals parameter $a$ such that the critical van der Waals temperature $kT_\text{c} = 8a/27b$ is somewhat greater than the clustering temperature at the critical density. This is true for our choice $a/\varepsilon \sigma^3=1$, for which $kT_\text{crit}/\varepsilon\approx
0.30$ for the pure van der Waals fluid and $kT_\text{clus}(\rho_\text{crit})/\varepsilon\approx 0.23$ in the absence of liquid formation (\textit{i.e.}~$a=0$).

As in subsection~\ref{subsect:equiline}, we can also calculate the clustering transition temperature, $T_\text{clus}(\rho)$, although we should note that where $T_\text{clus}(\rho)$ lies in the two-phase region, phase separation of course occurs instead of clustering. With liquid formation being favourable at high densities, however, the clustering transition curve does not simply increase in temperature as the density increases, but beyond $\rho\sigma^3 \approx 0.15$ instead gradually decreases, such that it reaches $\rho_\text{max}^\text{eql}$ at $kT/\varepsilon=0$, as shown in Fig.~\ref{fig-overall-phase-diag}.

The full phase diagram for $B/b=24$ is depicted in Fig.~\ref{fig-overall-phase-diag}. Reentrant phase behaviour is readily observed in this $T$-$\rho$ plane for $0.05 \lessapprox \rho\sigma^3 \lessapprox 0.5$. As the temperature decreases at constant density within this range, the system first undergoes a transition from a monomeric fluid to a two-phase liquid-vapour coexistence region, and then as the temperature is further decreased, to a fluid of icosahedral clusters.

\begin{figure}[tbp]
\centering
\includegraphics{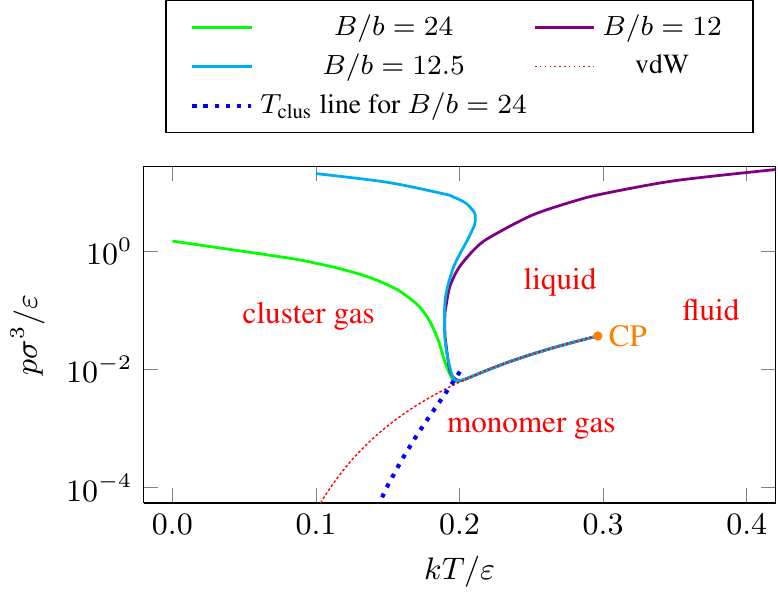}
\caption{The $p$-$T$ phase diagram for the self-assembling van der Waals fluid at different values of $B/b$. The critical point is labelled `CP'. The clustering temperature ($T_\text{clus}$) is also plotted for $B/b=24$ (thick dotted line) in the monomer-cluster region only.  $a/\varepsilon\sigma^3=1$, $b/\sigma^3=1$, $\sigma_\text{pw}=0.5$.
}\label{fig-PT-diags}
\end{figure}

At high temperatures, the self-assembling van der Waals fluid exhibits liquid-vapour coexistence behaviour exactly analogous to the pure van der Waals fluid, and the binodals of the self-assembling fluid follow those of the van der Waals fluid. This liquid-vapour phase separation is driven by the attractions between monomers. In this region, the phase boundary exhibits a positive gradient in the $p$-$T$ phase diagram (Fig.~\ref{fig-PT-diags}). The Clapeyron equation\cite{Atkins} relates this gradient to
\begin{equation}
\frac{\mathrm{d} p}{\mathrm{d} T} = \frac{\Delta_\text{trs} H}{T_\text{trs}\Delta_\text{trs} V} = \frac{\Delta_\text{trs} S}{\Delta_\text{trs} V},
\end{equation}
and since the entropy, enthalpy and volume changes are all negative for the vapour-liquid transition, the resulting slope is naturally positive.

By contrast, at sufficiently low temperatures, $T_\text{clus}(\rho)$ lies at lower density than the van der Waals binodal. Therefore, as the density increases,
the vapour forms clusters before it reaches the point at which it would otherwise have become phase separated. As clusters have no attractions between them, the vapour is now stable and does not phase separate.

As the density of this system approaches that for the maximum density of clusters, packing constraints become important. If the density is to move beyond $12\rho_\text{clus,\ max}$ and remain homogeneous, the system must form a monomer-cluster mixture with the proportion of monomers increasing with $\rho$. However, unlike in subsection~\ref{subsect:equiline},
the system now prefers to demix into a vapour of clusters and a monomeric liquid because of a lack of attractions between clusters and monomers.
Consequently, both the solubility of monomers in the cluster fluid and of clusters in the monomeric liquid is very low. For example, at $kT/\varepsilon=0.08$, $\rho_\text{mon}\sigma^3=3.44\times 10^{-7}$ at the lower-density binodal ($\rho\sigma^3 = 0.498$), and $\rho_\text{clus}\sigma^3=7.28\times 10^{-120}$ at the higher-density binodal ($\rho\sigma^3 = 0.955$).
This demixing is a density-driven transition, and in this regime the position of the lower-density binodal varies little with temperature. As the vapour phase is now lower in energy than the liquid due to the strong internal bonding in the clusters, the enthalpy change associated with vapour-liquid transition is now positive, while the volume change is of course still negative. Hence, the cluster-liquid transition has a negative slope in the $p$-$T$
diagram (Fig.~\ref{fig-PT-diags}).

The crossover between these two regimes occurs at temperatures near the point at which $T_\text{clus}(\rho)$ crosses the pure van der Waals binodal and leads to a rapid increase in the density of the lower-density binodal with decreasing
temperature until $12\rho_\text{clus,\ max}$ is approached. The width of this crossover regime reflects the width of the monomer-cluster equilibrium. The change in slope of the binodal occurs slightly above $T_\text{clus}$, as the vapour first begins to cluster. Initially, when monomers are still in the majority at the binodal, the transition is still driven by the attractions
between the monomers, and the binodal occurs when the monomer density reaches the density of the binodal for the pure van der Waals system. However, as clusters begin to predominate in the coexisting mixture, the phase transition is driven more by the lack of attractions between monomers and clusters, and occurs when the monomer density reaches a critical solubility.

These changes in the phase diagram are also reflected in the $p$-$v$ and $p$-$\mu$ diagrams (Fig.~\ref{fig-PVdiags}). For $0.23\,\varepsilon/k \lessapprox T < T_\text{crit}$, we observe van der Waals-like $p$-$v$ and $p$-$\mu$ curves. However, below this temperature, even though not yet in the region of reentrance, we can observe additional metastable closed curves in both $p$-$v$ and $p$-$\mu$ plots. As the temperature decreases, the cluster-rich loop (as seen at $kT/\varepsilon=0.21$) merges with the van der Waals curve and thus exhibits two additional spinodal points compared to the situation in the pure van der Waals case, due to a back-bending of the $p$-$v$ curve. This merging occurs when $T_\text{clus}(\rho)$ meets the spinodal density, as shown in Fig.~\ref{fig-overall-phase-diag}. Below this temperature, because the cluster predominates in the vapour, the vapour is no longer destabilised by the presence of attractions and instead only becomes unstable when the density goes beyond $12\rho_\text{clus,\ max}$.

To obtain spinodals, we numerically calculated derivatives obtained directly from $p$-$v$ curves. As already mentioned in subsection \ref{subsect:stabcrit}, the unusual sections of backbending in the $p$-$v$ isotherms (\textit{e.g.}~curve BC in Fig.~\ref{fig-PVdiags}(c)) are, although mechanically stable, compositionally unstable, and therefore the points at which $\partial p / \partial v = \infty$ are not `true' borders of (in)stability. The physically relevant spinodals are those of type A and B in the van der Waals-like region of the phase diagram, and then D instead of B in lower regions; these are shown in Fig.~\ref{fig-overall-phase-diag}.

For $B/b=24$, despite the reentrance, at all temperatures below $T_\text{crit}$ there is always a density range for which phase separation occurs.
We will now consider how the topology of the phase diagram can be modified by varying the parameters of the model.
In particular, we want to investigate whether the phase separation at low temperature can be removed, and if so, whether the phase diagram might exhibit a closed-loop similar to the nicotine-water mixture.

\begin{figure}[tbp]
\centering
\includegraphics{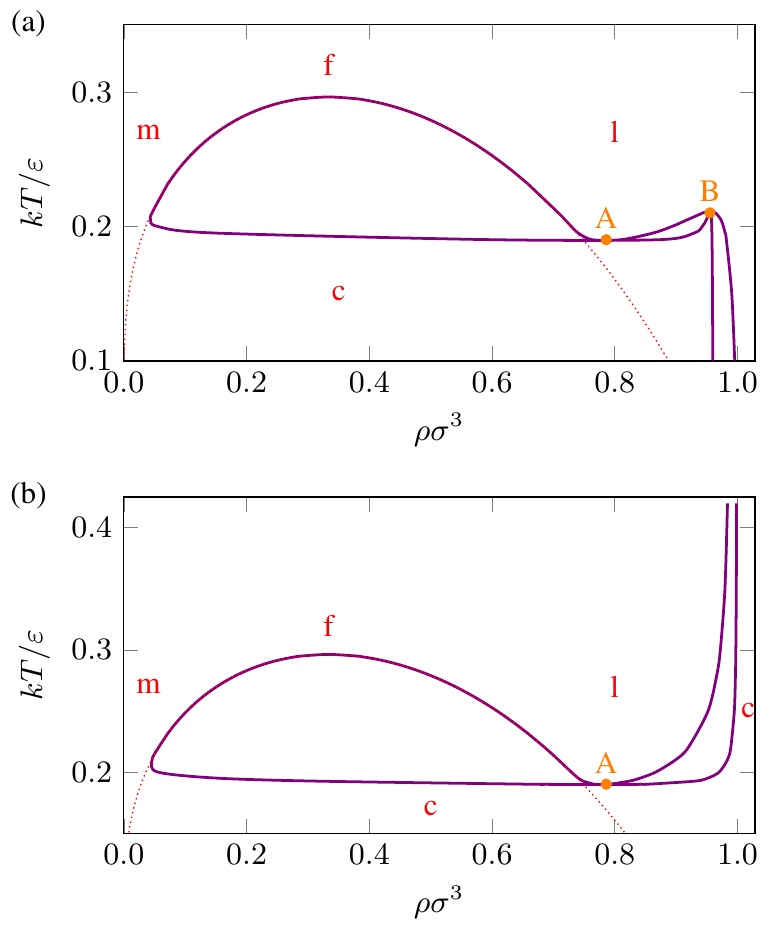}
\caption{Binodal lines for (a) $B/b=12.5$ and (b) $B/b=12$ are shown in violet, compared with the dotted red line representing the pure van der Waals binodal. The labels `c', `m', `l' and `f' refer to cluster gas, monomer gas, liquid and fluid states, respectively.  $a/\varepsilon\sigma^3=1$, $b/\sigma^3=1$, $\sigma_\text{pw}=0.5$.}\label{fig-phase-diag-B12}
\end{figure}

First, we investigate the effect of decreasing $B$, as this parameter determines $\rho_\text{clus,\ max}$ and
hence the onset of demixing in the low-temperature regime.
As $B/b$ decreases, the clusters are able to pack more efficiently and the lower-density binodal moves to higher density.
Initially, this does not lead to any change in the topology of the phase diagram.
However, the situation becomes more interesting as the binodal associated with the cluster fluid meets and crosses that associated with the monomeric liquid.
Fig.~\ref{fig-phase-diag-B12}(a) illustrates the type of phase diagram that results. We note that this scenario does not lead to a lower critical point; rather, at point A in the phase diagram, there are two coexisting fluids with the same density, but which differ significantly in other characteristics.
One fluid is a monomeric liquid with high entropy, while the other is a dense cluster fluid with low energy.
These two fluids do not evolve continuously into each other as point A is approached (as would need to be the case for there to be a critical point), but instead retain their separate identities.

As $\Delta V=0$ at point A in Fig.~\ref{fig-phase-diag-B12}, $\mathrm{d}p/\mathrm{d}T\to\infty$ and there
is a change in sign of the slope of the phase boundary in the $p$-$T$ phase
diagram. Initially, beyond this point, the coexisting cluster fluid is more
dense, and so $\mathrm{d}p/\mathrm{d}T$ is again positive. However, for $B/b>12$, the clusters
still pack less efficiently, and so at low temperature and densities beyond
$12\rho_\text{clus,\ max}$, there must again be a region of demixing
because of the lack of attractions between cluster and monomers.
The slope of the $p$-$T$ phase boundary corresponding to this demixing
must again be negative and so there must be a second point (B) where the
slope of the phase-boundary changes and where the coexisting fluids have
the same density. Curiously, for this phase diagram topology, there is a temperature range ($0.19\lessapprox kT/\varepsilon \lessapprox 0.21$ for $B/b=12.5$) where there are three fluid-fluid phase transitions as the system is pressurised.

Finally, at $B/b=12$, there must be a further change in the phase diagram topology, because at this point monomers no longer pack more efficiently than clusters, and at low temperature the cluster fluid will be more stable for any feasible density.
As $B/b$ approaches 12 from above, point B moves to higher and higher temperatures, and the density range for the cluster-fluid/monomeric liquid demixing becomes smaller and smaller until it disappears at $B/b=12$.
The resulting phase diagram is illustrated in Fig.~\ref{fig-phase-diag-B12}(b).

The second effect we wish to consider is changing the position of the liquid-vapour critical point with respect to the clustering transition.
For a given $b$, $T_\text{crit}$ is determined by the value of $a$, and $T_\text{clus}$ is determined by $\sigma_\text{pw}$.
In contrast to our model, for the simulated patchy particle systems, the two parameters are not independent, but instead the position of the critical point varies with the patch width (and the number of patches).\cite{Giacometti2010, Bianchi2006}  For example, it can be seen from Fig.~\ref{fig-yieldplot} that the temperature range associated with liquid-vapour coexistence
decreases as $\sigma_\text{pw}$ decreases. Here, we therefore choose to vary one of the parameters ($a$) whilst keeping the other one ($\sigma_\text{pw}$) fixed.

\begin{figure}[tbp]
\centering
\includegraphics{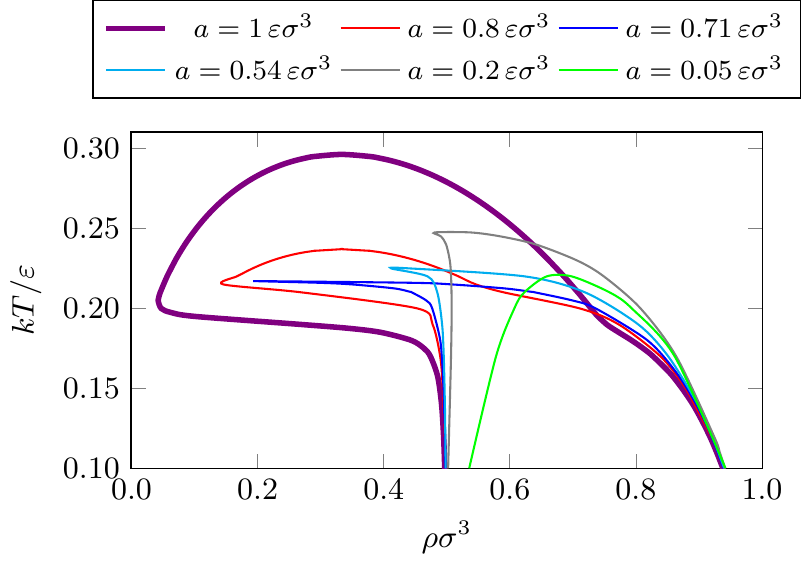}
\caption{A $T$-$\rho$ phase diagram for the self-assembling van der Waals fluid with a selection of values for the van der Waals attractive parameter $a$. $b/\sigma^3=1$, $b/B=24$, $\sigma_\text{pw}=0.5$.}\label{fig-phase-diag-changeA}
\end{figure}

The results for a variety of values of $a$ are shown in Fig.~\ref{fig-phase-diag-changeA}. When the van der Waals critical temperature is above the clustering temperature (for example, at $a=1\,\varepsilon\sigma^3$ and $a=0.8\,\varepsilon\sigma^3$, which correspond to critical temperatures of $kT_\text{crit}/\varepsilon=0.296$ and $kT_\text{crit}/\varepsilon=0.237$, respectively), van der Waals behaviour is observed at high temperatures. When the critical temperature descends below the clustering temperature, we nevertheless continue to observe a bending of the binodal curve to lower densities in the region where liquid-vapour coexistence previously took place.
However, this feature is not some remnant of the liquid-vapour transition, but is instead associated with demixing in the vicinity of $T_\text{clus}(\rho)$.
At $T_\text{clus}$, there is an equal probability of particles being monomeric or as part of clusters; however, if $\rho$ and $a$ are sufficiently large, the system will not form a homogeneous mixture, but will instead demix.
The narrowness in temperature of this feature is associated with the sharpness with which the chemical equilibrium shifts between being monomer dominated to being cluster dominated as the temperature decreases.
As $a$ decreases, there is a progressively smaller demixing driving force and so the size of this feature decreases.
Note that these changes lead to a non-monotonic dependence of $T_\text{crit}$ on $a$, because $T_\text{crit}$ initially increases as $\rho_\text{crit}$ increases, mirroring the dependence of $T_\text{clus}$ on $\rho$ (Fig.~\ref{fig-cluster-gas}).
As $a$ becomes smaller still, the clusters and monomers begin to be able to mix at high temperatures because less entropy of mixing is required to overcome the enthalpic mixing incompatibility.
Hence, $T_\text{crit}$ again begins to decrease.  However, even at a small value of $a$, \textit{e.g.}~$a=0.05\,\varepsilon\sigma^3$ in Fig.~\ref{fig-phase-diag-changeA}, there is still a substantial region of demixing. As $a$ tends to zero, the demixing regime disappears, recovering the monomer-cluster equilibrium studied in section~\ref{sec:mon-clus}.

\begin{figure}[tbp]
\centering
\includegraphics{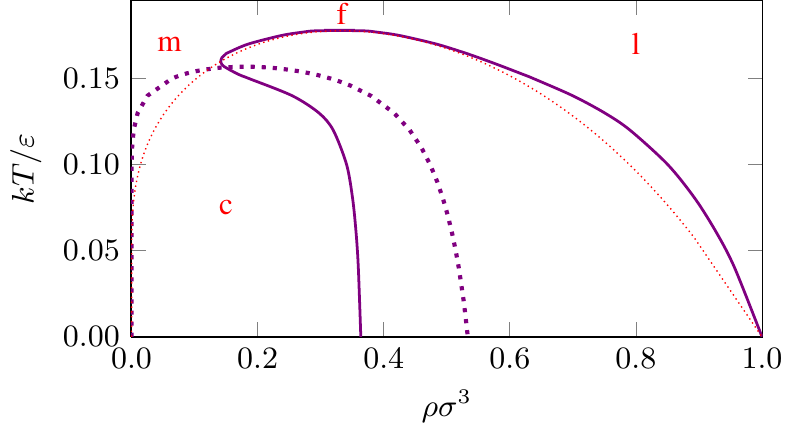}
\caption{A $T$-$\rho$ phase diagram for the self-assembling van der Waals fluid forming tetrahedral clusters. The solid line is the binodal curve, whilst the thin dotted line represents the pure van der Waals system binodal. The thick dashed line is the clustering temperature, $T_\text{clus}(\rho)$, for this system. The labels `c', `m', `l' and `f' refer to cluster gas, monomer gas, liquid and fluid states, respectively. $a/\varepsilon\sigma^3=0.6$, $B/b=11$, $b/\sigma^3=1$, $\sigma_\text{pw}=0.5$.}\label{fig-phase-diag-tetrahedra}
\end{figure}

So far, we have exclusively considered particles that can self-assemble into icosahedral clusters, but our approach can easily be generalised to the formation of other types of cluster.
Here, we illustrate this by calculating the phase diagram for particles able to form tetrahedra.\cite{Wilber2009}
The requisite changes in the model from the icosahedral system discussed above are as follows. The moment of inertia of a tetrahedron about the principal axes is $I=4m\sigma^2$, and the mass of the cluster is $4m$. The symmetry number of a tetrahedron is 12, and that of an individual monomer is 3. The ground state energy of the tetrahedron is $-6\varepsilon$. Finally, the circumradius of a tetrahedron with edge length $a$ is $r=\sqrt{3/8}\,a$,\cite{CRC-tables} and so the physical value of $B/b$ which we have used is $B/b=11$. This gives a maximum particle density associated with clusters of $4\rho_\text{clus,\ max}\sigma^3=4/11\approx 0.36$. We have assumed that tetrahedra will pack approximately like spheres; this approximation is not so far-fetched as it might seem at first glance, and recent research confirms that the packing fractions are very similar.\cite{Torquato2009} We also scale the value of the parameter $a$ by 3/5 compared to that for the icosahedron in order to take into account that there are only three rather than five patches that give rise to the interactions that drive the liquid-vapour phase separation.
The phase diagram of this system is depicted in Fig.~\ref{fig-phase-diag-tetrahedra}. The behaviour of the system is similar to the icosahedral system, with the main differences being easy to rationalise. For example, the clustering temperature is significantly lower due to the lower energetic driving force to form clusters. Similarly, the crossover between the liquid-vapour and demixing regimes occurs over a wider temperature range due to the greater width of the monomer-cluster equilibrium due to the smaller size of the clusters.

\section{Discussion and conclusions}\label{sec:discussion}
We have calculated the complete phase diagram for a system exhibiting competition between phase separation and self-assembly into monodisperse clusters.
For nearly all of the parameterisations of the model that we have considered, decreasing the temperature at constant density results, for a limited range of density, in the transition from a gas of monomers, to a liquid-gas phase-separated mixture, and then to a gas of clusters. Therefore, the existence of reentrant phase behaviour, driven by the lower energy of clusters, is clear within this model and is a robust feature of such systems.
In this aspect, the model agrees with our group's simulation results for patchy colloidal particles,\cite{Wilber2007,Wilber2009} and the Janus particle simulations of Sciortino and co-workers,\cite{Sciortino2009,Sciortino2010} both of which showed evidence of reentrant phase behaviour.

In particular, a comparison of the final phase diagram with the yield plot shown in Fig.~\ref{fig-yieldplot} shows that decreasing the temperature at a constant density of $\rho\sigma^3=0.15$ and a patch width of $\sigma_\text{pw}=0.5$ involves a reentrant phase transition in both cases. Even though we have not chosen to match the parameter $a$ precisely to the icosahedral system studied in simulations, there is surprisingly good agreement not only in the underlying physical behaviour, but also in the numerical values.

While the phase behaviour of the fluid at low temperature is not yet fully clear in
the simulated systems, our model allows the complete phase diagram to be calculated.
We have established that phase separation persists at low temperatures within the framework of our model. The driving force for it, however, is not the attractions between monomers as in the pure van der Waals fluid, but rather cluster-liquid immiscibility. For this reason, reentrance is observed over only a limited range of density in all physically relevant phase diagrams; at higher densities, demixing between the cluster and the liquid states ensures continued phase separation.

In our model, the change in topology of the phase diagram as the model parameters are varied is easy to calculate. The free volume parameter ratio $B/b$ changes the maximum density of the cluster, and therefore affects the position of the lower-density binodal of the $T$-$\rho$ phase diagram in the demixing regime. The change in the van der Waals parameter $a$, on the other hand, changes the van der Waals critical temperature, and, provided that it is greater than the clustering temperature, affects the temperature range of the phase-separated region.

Of course, in our theoretical model, there is a significant number of approximations, the effect of which we should try to understand, in particular in relation to the comparison of our results with phase diagrams obtained from simulations.
Firstly, our focus has just been on the fluid behaviour of these self-assembling systems, and, unlike in simulations, we can explore the low temperature form of the fluid phase behaviour without having to worry about other phases obscuring the fundamental behaviour.
However, crystallisation (of both monomers and clusters) is likely to have a major effect on the overall form of the phase diagram. There are precedents for such cluster crystals; for example, icosahedral virus particles (monodisperse aggregates of virus capsid proteins, perhaps with a nucleic acid genome inside) are often observed to form crystals when they are densely packed in cells.\cite{Doye2006}

\begin{figure}[tbp]
\centering
\includegraphics{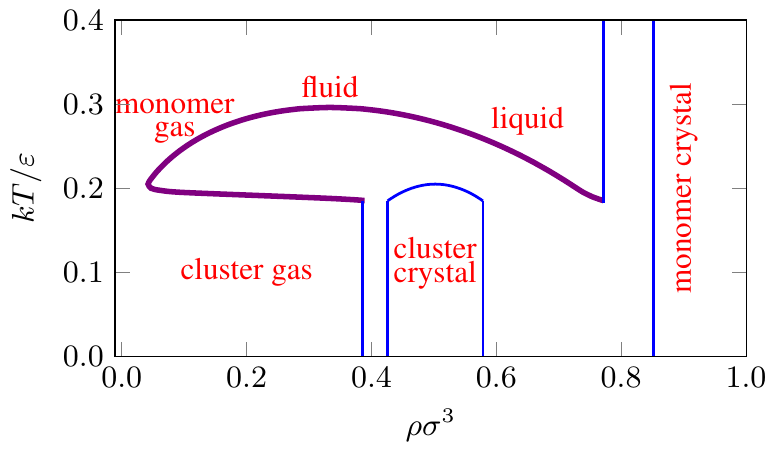}
\caption{A schematic $T$-$\rho$ phase diagram to illustrate the potential effects of crystallisation on the phase diagram in Fig.~\ref{fig-overall-phase-diag}.}\label{fig-overall-HS-crystal}
\end{figure}

We can construct a schematic phase diagram illustrating the potential effects of crystallisation on it by assuming hard-sphere crystallisation of both components.
This approximation is probably quite reasonable for the clusters, as all the attractive patches are involved in the internal bonding of the cluster, but probably less so for the monomers, where the effect of the patches on the crystal stability and form is less clear.
For hard spheres, the fluid-crystal coexistence limits occur at packing fractions $\phi=0.494$ and $\phi=0.545$,\cite{Anderson2002} and by associating the maximum densities of the fluids with the packing fraction in random close packing, $\phi=0.64$,\cite{Torquato2000} we can obtain estimates for the densities of the fluid-crystal binodals.
Further, the maximum hard-sphere crystalline packing fraction, $\phi=0.74$, provides a means for obtaining an upper density limit for the stability of the cluster crystal.
Finally, the clustering temperature provides an upper temperature stability limit for the cluster crystal. The resulting schematic phase diagram is shown in Fig.~\ref{fig-overall-HS-crystal}. Note that this phase diagram has a second triple point associated with the disappearance of the cluster crystal phase, as well as the usual one associated with the disappearance of the liquid phase. We have not attempted to consider the effect of other possible phases on the phase diagram, such as the lamellar phase observed for Janus particles\cite{Sciortino2009,Sciortino2010}
and the two-dimensional crystalline lamellae\cite{WilliamsonUnpublished} seen in simulations of octahedron-forming patchy particles.\cite{Wilber2009}

Secondly, there are also a series of approximations associated with our description of the fluid.
For example, the free volume approximation\cite{Levin1996} is not flexible enough to fully capture both the high- and the low-density behaviour (we choose $B$ and $b$ values appropriate to the high-density limit) and the form of the available volume term differs from that for the exact hard-sphere second virial coefficient for interactions between monomers and clusters. There is also an obvious problem when the density of clusters approaches the maximum allowed, as the free volume approximation predicts that the translational partition function of both clusters and monomers should tend to zero. However, because of the large differences in size between monomers and icosahedral clusters, even when the clusters are close-packed, there is still space available for the monomers in the gaps between the clusters. In other words, the translational partition function of the monomers should not tend to zero as the translational partition function of the clusters does so. However, such correlation effects associated with the positions of the particles are not so straightforward to take into account.

In our model, we also effectively assume that the clusters pack, in essence, like hard spheres, without any change of shape or compression, regardless of the overall density of the system. This assumption, which is responsible for the near vertical slope of the lower-density binodal in the low-temperature demixing regime, may be reasonable for monodisperse clusters that have a well-defined shape and are held together by specific bonds (although for our patchy particle potential, the clusters will
be slightly compressible). However, this assumed cluster incompressibility may be a factor in the mismatch in the detailed shape of our phase diagrams compared to those obtained in the simulations by Sciortino and co-workers,\cite{Sciortino2009,Sciortino2010} where the shapes of clusters are much more deformable, as well as not being monodisperse. In their computed phase diagrams, they observe the binodals moving to higher densities with decreasing temperature considerably more gradually than we do.

Another approximation we have made is that we can model the monomeric liquid and vapour phases as a van der Waals fluid. In this model, the critical density has a fixed value (depending only on the free volume parameter $b$), whereas we know that in patchy particle colloids, the critical point varies with both the patch width and the patch number.\cite{Bianchi2006, Giacometti2010} Nevertheless, this deficiency is probably not so important for the qualitative shape of the phase diagram, especially as we are most concerned in this paper with the behaviour at lower temperature associated with the onset of clustering.

Finally, we have assumed that clusters form either completely or not at all, \textit{i.e.}~only a single size of clusters can be formed. Although this is an approximation, the probability of observing partly-formed clusters \emph{at equilibrium} is very small, and so the effect on the phase diagrams will be negligible.\cite{Wilber2007}

Although these approximations will affect the quantitative comparison between the phase diagrams for the current model and those computed in patchy particle simulations,\cite{Wilber2007,Wilber2009} and hopefully at some future point those found experimentally for self-assembled patchy colloids, we believe that the fundamental insights the current model gives will help to provide a theoretical framework for understanding the fluid phase behaviour of these kinds of system.

\begin{acknowledgments}
We would like to thank the Engineering and Physical Sciences Research Council, the Royal Society, the Spanish Ministry of Education and Science and the European Regional Development Fund (grant FIS2007-66823-C02-02) for financial support. J.~Carrete wishes to thank the Spanish Ministry of Education and Science for an FPU grant.
\end{acknowledgments}


%

\end{document}